\documentclass[reprint,
groupedaddress,
 prl,amsmath,amssymb,
 aps]{revtex4-2}
 
\usepackage[normalem]{ulem}
\usepackage{appendix}
\usepackage{blindtext}
\usepackage{multirow}    
\usepackage{graphicx}
\usepackage{mathtools}
\usepackage{dcolumn}
\usepackage{amsthm}
\usepackage{qcircuit}
\usepackage{amsmath}
\usepackage{amsfonts}
\usepackage{amssymb}
\usepackage{ctable}
\usepackage{bm}
\usepackage{bbm}
\usepackage{color}
\usepackage{placeins}
\usepackage{afterpage}
\usepackage{comment}
\usepackage{wasysym}
\usepackage{amsmath}
\usepackage{bibunits}
\defaultbibliographystyle{plain}
\usepackage{hyperref}
\hypersetup{
    colorlinks=true,
    linkcolor=red,
    urlcolor=red,  
}

\makeatletter
\newcommand{\vast}{\bBigg@{4}}
\newcommand{\Vast}{\bBigg@{5}}
\makeatother

\makeatletter
\def\maketitle{
	\@author@finish
	\title@column\titleblock@produce
	\suppressfloats[t]}
\makeatother

\theoremstyle{definition}

\usepackage{braket}

\begin{document}

\title{Tailoring fusion-based error correction for high thresholds to biased fusion failures}

\author{Kaavya Sahay}
\author{Jahan Claes}
\author{Shruti Puri}
\affiliation{Department of Applied Physics, Yale University, New Haven, Connecticut 06511, USA\\
Yale Quantum Institute, Yale University, New Haven, Connecticut 06511, USA}

\begin{abstract}

We introduce fault-tolerant (FT) architectures for error correction with the XZZX cluster state based on performing measurements of two-qubit Pauli operators $Z\otimes Z$ and $X\otimes X$, or {\it fusions}, on a collection of few-body entangled resource states. 
Our construction is tailored to be effective against noise that predominantly causes faulty $X\otimes X$ measurements during fusions. This feature offers practical advantage in linear optical quantum computing with dual-rail photonic qubits, where failed fusions only erase $X\otimes X$ measurement outcomes.
By applying our construction to this platform, we find a record high FT threshold to fusion failures exceeding $25\%$ in the experimentally relevant regime of non-zero loss rate per photon, considerably simplifying hardware requirements.

\end{abstract}
\date{\today}

\maketitle

\begin{bibunit}[apsrev_longbib]

{\it Introduction---} Fault-tolerant (FT) error correction enables arbitrary suppression of errors as long as the error rate is below a constant threshold, making scalable quantum computation possible. It is important to take into consideration the underlying physical operations available when designing a FT architecture. For example, if the entangling operations are inherently probabilistic or if the noise in these operations destroys the qubits, then the framework of FT {\it measurement-based} error correction (MBEC) is more natural~\cite{menicucci2006universal,bourassa2021blueprint,bartolucci2021fusion}. MBEC is implemented using a {\it cluster state}~\cite{briegel2001persistent,raussendorf2001one,raussendorf2002one,raussendorf2003measurement,briegel2009measurement}, which is a many-body entangled state and may be obtained from a stabilizer error correcting code using a process called {\it foliation}~\cite{raussendorf2006fault,raussendorf2007fault, bolt2016foliated,brown2020universal}. Outcomes of {single-qubit} measurements performed on the cluster state are used to reconstruct the underlying stabilizers and correct errors~\cite{raussendorf2006fault,raussendorf2007topological}. { {The measured qubits are removed from the entangled state, allowing considerable flexibility in how the measurements are realized in hardware. For example, it is possible for the measurement to be destructive and erase the measured qubit.}} The most well known cluster state is the Raussendorf-Harrington-Goyal (RHG) cluster state~\cite{raussendorf2005long,raussendorf2006fault,raussendorf2007topological} which is a foliation of the standard surface code~\cite{kitaev1997quantum}. Recently, the XZZX cluster state was introduced~\cite{claes2022tailored}, which is a foliation of the XZZX surface code~\cite{wen2003quantum,ataides2021xzzx}.

In the most common MBEC framework, the cluster state is generated using a set of commuting two-qubit entangling gates. Alternatively, one could start with a collection of few-body entangled states and then stitch them together into a many-body entangled cluster state using measurements of two-qubit Pauli operators $X\otimes X$ and $Z\otimes Z$, also called {\it fusions} or {\it Bell measurements}, { {which may be implemented destructively~\cite{browne2005resource}.}} This approach has been referred to as {\it fusion-based error correction} (FBEC)~\cite{bartolucci2021fusion} and is a more natural choice for architectures where 
 {{high-fidelity fusions are native to the hardware}} like discrete variable photonic qubits~\cite{browne2005resource}, {{continuous variable qubits~\cite{fukui2018high,fukui2019high}}}, and Majorana-based qubits~\cite{chao2020optimization}. {{The FBEC framework has been studied for error correction with the RHG cluster state ~\cite{bartolucci2021fusion} and, more recently, with the foliated floquet color code~\cite{paesani2022high}. }}

In this paper, we introduce fusion-based architecture for error correction with the XZZX cluster state. We present two constructions, one based on using a collection of 4-body entangled resource states and the other based on using a set of 6-body entangled resource states. Importantly, both the constructions offer practical advantage when noise in the fusion circuit is biased so that the $Z\otimes Z$ measurements are much more reliable than $X\otimes X$ measurements. This is because errors introduced in the cluster states due to faulty $X\otimes X$ measurements, referred to as biased fusion failures, give rise to a two-dimensional system symmetry~\cite{brown2022conservation} which considerably simplifies the decoding problem, leading to a substantial improvement in the threshold to biased fusion noise.

Our construction is motivated by dual-rail qubits in linear optics~\cite{nielsen2004optical,nielsen2005fault,gilchrist2007efficient,kok2007linear}, which is the most widely studied platform in the framework of FBEC~\cite{browne2005resource,li2015resource,bartolucci2021fusion}. Linear-optic fusions on dual rail qubits are inherently probabilistic. The simplest fusion circuit fails with probability $1/2$~\cite{browne2005resource}. 
The failure probability can be reduced to $1/2^n$ using an ancillary $(2^n-2)$-photon entangled state~\cite{grice2011arbitrarily}, although for the special case of {$1/4$ failure probability}, $4$ unentangled photons are {also} sufficient~\cite{ewert20143}. Notably, when a fusion attempt fails, the $X\otimes X$ information is completely erased but $Z\otimes Z$ can still be recovered~\cite{browne2005resource,bartolucci2021fusion}. The architecture proposed here leverages this biased noise structure to achieve record-high thresholds to fusion failures. With numerical simulations of the fusion-based XZZX cluster state with photonic dual-rail qubits and entangled ancillae, we find a threshold to fusion failures exceeding $25\%$ in the experimentally relevant regime of non-zero loss rate per photon. This is the highest known threshold to fusion failures in linear optics without additional encodings on the photonic state{{~\cite{li2010fault,auger2018fault,bartolucci2021fusion,paesani2022high}}}, and for the first time allows scalable FBEC using an ancilla of only two entangled photons or four unentangled photons.

{\it The XZZX cluster state--} We start with a review of the XZZX cluster state. It is a specific instance of a generalized cluster state~\cite{claes2022tailored}, a stabilizer state defined on a decorated graph $G=(V,E)$ with two types of vertices $V=\mathcal{X}\sqcup\mathcal{Z}$. Each vertex represents a qubit of the cluster state; we refer to $v\in\mathcal X$ as $X$-type qubits and denote them by $\newmoon$, and we refer to $v\in \mathcal Z$ as $Z$-type qubits and denote them by $\bigcirc$. The $N$-qubit generalized cluster state is the $+1$ eigenstate of $N$ mutually commuting stabilizers, one centered at each qubit $v\in V$, given by
\begin{equation}
 \left\{\begin{aligned}X_v\prod_{\substack{(v,w)\in E\\w\in\mathcal{Z}}}X_w\prod_{\substack{(v,u)\in E\\u\in\mathcal{X}}}Z_u, & \quad v\in \mathcal{X}\\
 Z_v\prod_{\substack{(v,w)\in E\\w\in\mathcal{Z}}}X_w\prod_{\substack{(v,u)\in E\\u\in\mathcal{X}}}Z_u, & \quad v\in \mathcal{Z} 
 \end{aligned}\right. .\label{eq:GeneralizedClusterStabilizers}
\end{equation}
Essentially, for each $v\in \mathcal{X}$ we have a stabilizer given by the product of $X_v$ on that qubit, and some combination of $X$ and $Z$ operators on the neighboring qubits. Similarly, for each $v\in \mathcal{Z}$ we have a stabilizer given by the product of $Z_v$ on that qubit, and some combination of $X$ and $Z$ operators on the neighbors. These neighboring $X$ and $Z$ operators are chosen to ensure all stabilizers commute.

The XZZX cluster state is defined on a periodic 3D graph, a unit cell of which is shown in Fig.~\ref{fig:XZZXClusterStateRefresher}(a), along with stabilizers centered at an $X$-type qubit and $Z$-type qubit. Note that there are no edges between $Z$-type qubits. The XZZX cluster state has the same geometry as the RHG cluster state~\cite{raussendorf2005long,raussendorf2006fault}, and can be obtained from it by local Clifford operations. Taking the product of the stabilizers centered on the faces of a unit cell gives the cell stabilizer shown in Fig.~\ref{fig:XZZXClusterStateRefresher}(b), which is used to correct errors. Performing a measurement-based computation using the XZZX cluster state involves measuring all $Z$-type qubits in the $Z$ basis and all $X$-type qubits in the $X$ basis. Once we have measured all qubits in their respective bases, we use the cell stabilizers to check for errors in our measurement outcomes. A $Z$ error on an $X$-type qubit or an $X$ error on a $Z$-type qubit causes {the qubit's two} neighboring cell stabilizers to flip to $(-1)$, as shown in Fig.~\ref{fig:XZZXClusterStateRefresher}(c). Importantly, we note that $Z$ errors on $X$-type qubits only cause pairs of defects that are restricted to 2D planes. This 2D system symmetry simplifies the decoding problem{---for example,} a matching decoder only needs to match defects in 2D---and ultimately leads to higher thresholds for $Z$-biased noise~\cite{claes2022tailored,brown2022conservation}.

\begin{figure}
    \centering
    \includegraphics[width=\columnwidth]{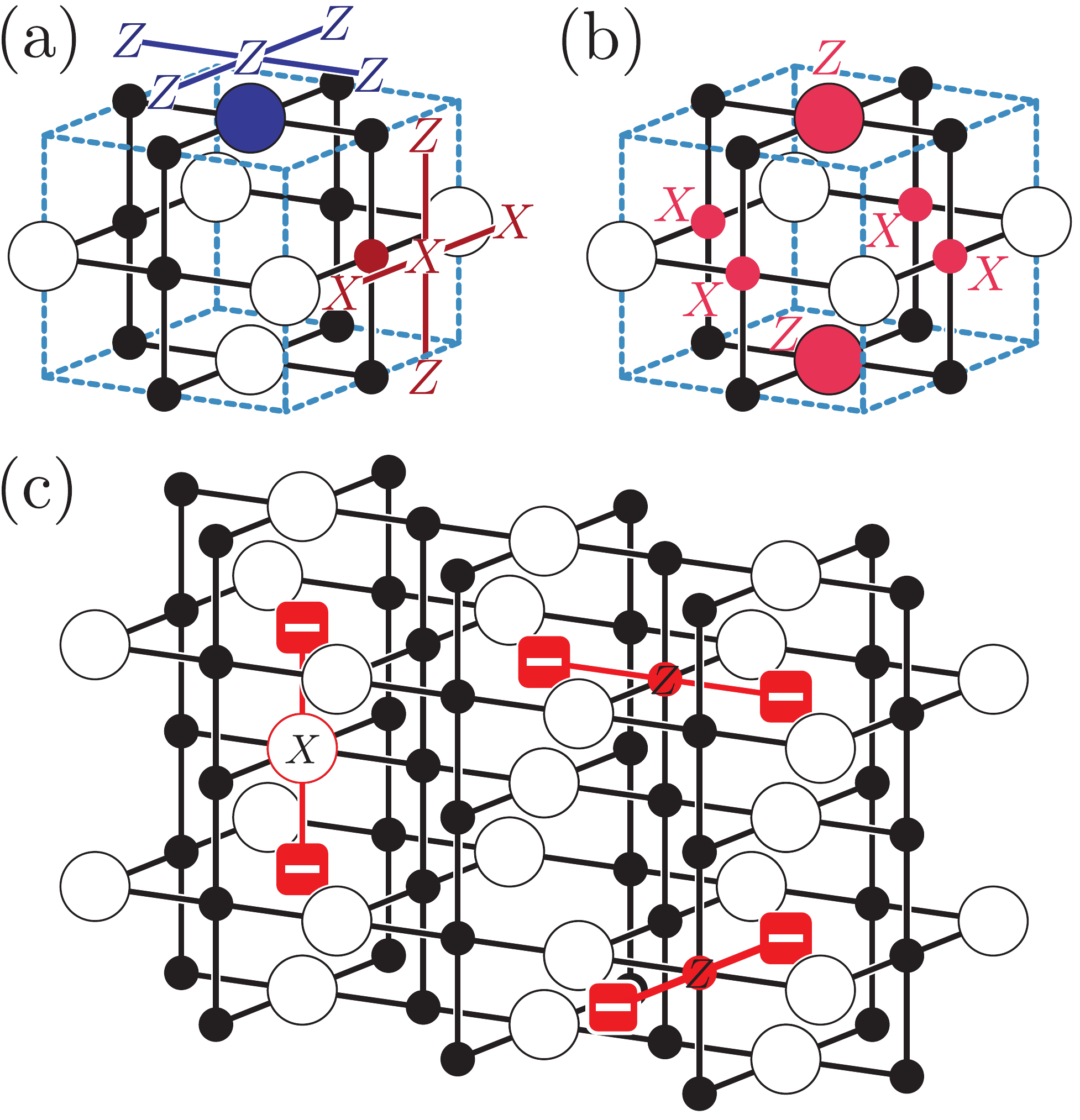}
    \caption{(a) A unit cell of the XZZX cluster state, with two examples of stabilizers centered on the faces of the cell as given by Eq.~\eqref{eq:GeneralizedClusterStabilizers}. (b) If we multiply the stabilizers centered at all faces of a unit cell, we get the cell stabilizer shown. (c) The value of the cell stabilizer allows us to detect errors. $X$ errors on $Z$-type qubits or $Z$ errors on $X$-type qubits cause the value of the neighboring cell stabilizers to flip. Importantly, $Z$ errors cause pairs of defects restricted to 2D planes, which allows for more effective decoding of $Z$ errors.}
    \label{fig:XZZXClusterStateRefresher}
\end{figure}

The XZZX cluster state may be prepared using controlled-not and controlled-phase gates, and the robustness of this cluster state to gate noise has been studied previously~\cite{claes2022tailored}. In this work we consider the alternative approach of preparing this state by fusing copies of few-body entangled resource states, which is the standard approach for realizing cluster states in photonic dual-rail platforms~\cite{nielsen2004optical,browne2005resource,herrera2010photonic,bartolucci2021fusion}.  We will consider two schemes for the 
XZZX cluster state, which expand on schemes introduced in Ref.~\cite{bartolucci2021fusion} for the RHG cluster state. The important distinction is that in both of our schemes biased fusion failures {create pairs of defects restricted to 2D planes}, leading to improved thresholds.

\begin{figure}
    \centering
    \includegraphics[width=\columnwidth]{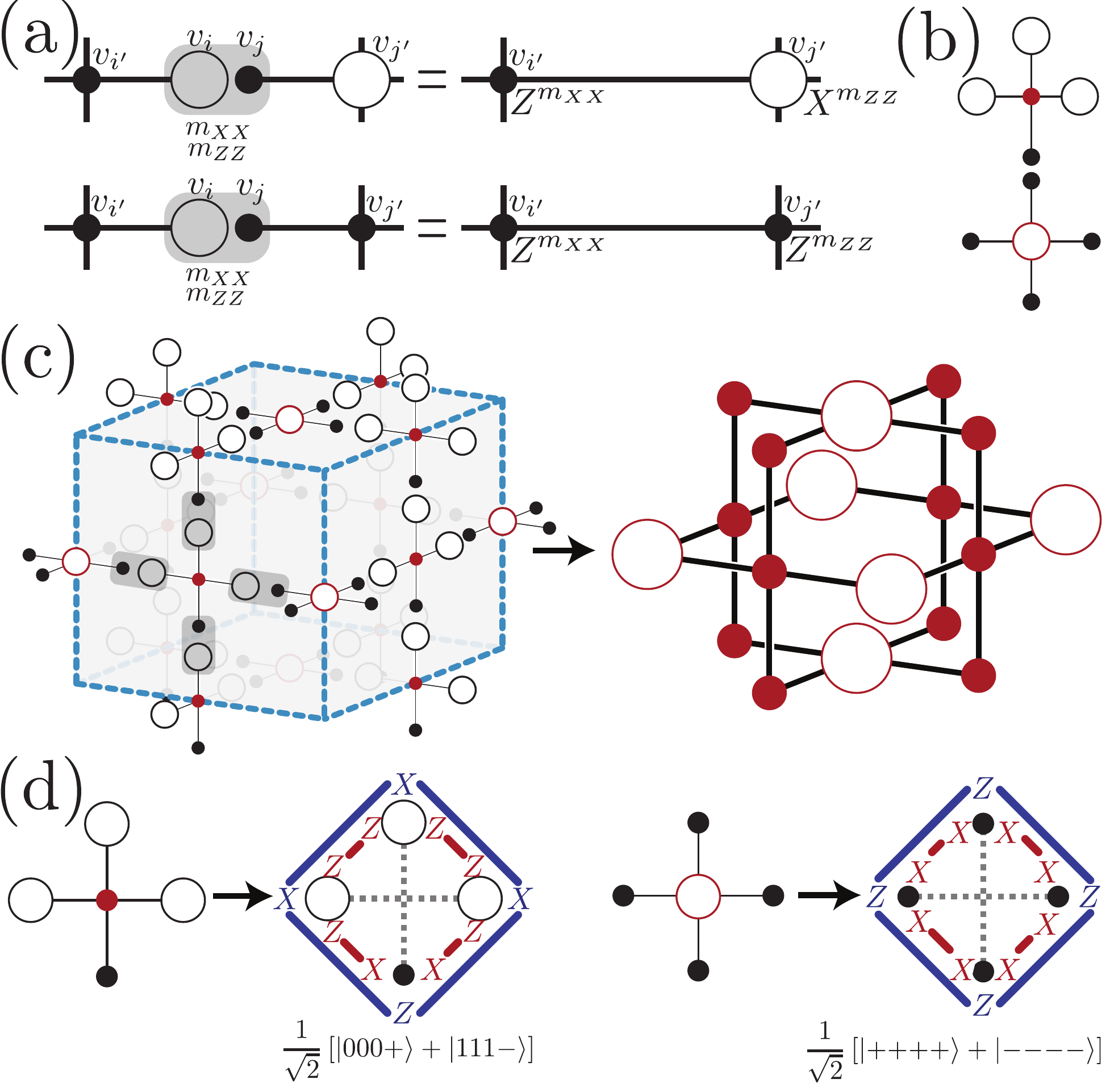}
    \caption{The 4-star construction. (a) Performing a fusion measurement of $X_i\otimes X_j$ and $Z_i\otimes  Z_j$ on a dangling pair of $X$- and $Z$-type qubits removes them from the cluster but forms an edge between their neighbors. (b) The two five-qubit cluster states we use in our construction. (c) The arrangement of five-qubit cluster states we use to build the XZZX cluster state. (d) Because the qubits of the resulting cluster state will be measured in the $X$/$Z$ basis, and because these measurements commute with the fusion measurements, we can measure the center qubits before performing the fusions. Equivalently, we can instead start with the simpler four-qubit resource states that result from measuring the center qubit as $(+1)$ in the $X$/$Z$ basis. In this case, the central qubits making up the cluster state are never physically realized, but exist as virtual qubits whose measurement outcomes are tracked in software.}
    \label{fig:TypeIIFusionClusterState}
\end{figure}

{\it Construction from 4-star resource states---} In the following discussion we introduce the principle underlying our construction. Consider a cluster state defined on a graph $G=(V,E)$. Let $G$ have a $Z$-type qubit at a degree-1 vertex $v_i\in\mathcal{Z}$ with an edge to a qubit at $v_{i'}$, and an $X$-type qubit at degree-1 vertex $v_j\in\mathcal{X}$ with an edge to a qubit at $v_{j'}\neq v_{i'}$. We will refer to the qubits on degree-1 vertices as {\it dangling qubits}. As shown in Fig.~\ref{fig:TypeIIFusionClusterState}(a), performing $X_i\otimes X_j$ and $Z_i\otimes Z_j$ measurements on the pair of dangling qubits disentangles them from the rest of the system {while entangling their neighbors}, removing vertices $v_i,\;v_j$ and edges $(v_i,v_{i'}),\; (v_j,v_{j'})$ and adding a new edge $(v_{i'},v_{j'})$. Consequently, the stabilizers centered at $v_i$ and $v_j$ are removed and we obtain two new stabilizers centered at $v_{i'}$ and $v_{j'}$ defined according to Eq.~\eqref{eq:GeneralizedClusterStabilizers}. To ensure that the new cluster state is the $+1$ eigenstate of these new stabilizers, a Pauli correction is applied to the qubits at $v_{i'}$ and $v_{j'}$ according to the outcomes of the $X_i\otimes X_j$ measurement {{($m_{XX}=$ 0 or 1)}} and $Z_i\otimes Z_j$ measurement {{($m_{ZZ}=$ 0 or 1)}}. If $v_{i'}\in\mathcal{X}$ and $v_{j'}\in\mathcal{Z}$, the correction is $Z_{i'}^{m_{XX}}\otimes X_{j'}^{m_{ZZ}}$, while if $v_{i'},v_{j'}\in\mathcal{X}$, the correction is $Z_{i'}^{m_{XX}}\otimes Z_{j'}^{m_{ZZ}}$. It is not necessary to physically apply these Pauli corrections and instead they may just be tracked in software. Observe that in case of unreliable $X_i\otimes X_j\; (Z_i\otimes Z_j)$ measurements, we cannot correctly determine the proper Pauli correction on $v_{i'}$ ($v_{j'}$) which results in an effective error on that qubit. In fact, a complete erasure of $X_i\otimes X_j$ measurement outcome that arises due to fusion failure in linear-optics is equivalent to applying $I$ or $Z$ to the $X$-type qubit at $v_{i'}$ with $50\%$ probability~\cite{supplement}. This type of $Z$-biased error at a known location in the cluster state ($v_{i'}$), marked by the location of the failed ($X_i\otimes X_j$) measurement, is classified as $Z$-{\it biased fusion failure}. 

With the above discussion in mind we introduce the two five-qubit cluster states shown in Fig.~\ref{fig:TypeIIFusionClusterState}(b) with stabilizers defined according to Eq.~\eqref{eq:GeneralizedClusterStabilizers}. One has a $Z$-type qubit at the center and the other has a $X$-type qubit. The center qubits are marked in red to indicate that these will eventually form the desired XZZX cluster state. The $Z$-centered and $X$-centered states are placed at the location of $Z$- and $X$-type qubits respectively in the desired cluster state, as shown in Fig.~\ref{fig:TypeIIFusionClusterState}(c). The arrangement of the 5-body states ensures that neighboring dangling qubits are always opposite types; we can thus fuse the neighboring dangling qubits according to Fig.~\ref{fig:TypeIIFusionClusterState}(a). The fused qubits are removed from the cluster and new bonds appear between the red qubits, resulting in the desired XZZX cluster state. Finally, the cluster state qubits can be measured in the appropriate basis described in the previous section for error correction. Note that each center qubit is entangled into the final cluster state after four fusion measurements on its neighboring dangling qubits; consequently, four Pauli corrections need to be accounted for on this qubit. This accounting may be done in software by simply re-interpreting the outcome of the final measurement of cluster state qubits. This is because a $X$ ($Z$) measurement of the $X$- ($Z$-) type qubit in the cluster state after a Pauli $Z$ ($X$) is applied to it is equivalent to a $X$ ($Z$) measurement followed by a classical flip of the measurement outcome {{$0\rightarrow 1,\; 1\rightarrow 0$}}.

It is possible to simplify the initial 5-qubit cluster states to 4-qubit resource states by noting that measuring the qubits comprising the final cluster state commutes with fusion measurements, as these measurements are performed on different qubits and Pauli corrections due to fusions can simply be accounted for in software. That is, we can equally well measure the center qubits which would form the XZZX cluster state before performing the fusion measurements. Once the fusions are realized, the outcomes of the prior center-qubit measurements can be flipped conditional on the fusion outcomes. Measuring the center $X$- and $Z$-type qubits of the 5-qubit states in the $X$ and $Z$ basis respectively leaves us with the simpler 4-star resource states shown in Fig.~\ref{fig:TypeIIFusionClusterState}(d). Thus we can directly start with these resource states assuming that the center qubits have been measured in the $X$/$Z$ basis with measurement outcome {{$0$}}. In this approach, the central qubit acts like a virtual qubit. It is never physically realized and never physically measured, and its effective measurement outcome is entirely determined by the Pauli corrections that are tracked in software~\cite{bartolucci2021fusion}.

{\it Construction from 6-ring resource states---} While the previous construction relied on fusing an $X$-type qubit with a $Z$-type qubit, this second construction relies on fusing two qubits of the same type. While our construction is reminiscent of the approach in ~\cite{bartolucci2021fusion}, we provide an alternate intuitive interpretation of why it works.

Consider a cluster state defined on a graph $G=(V,E)$ with $Z\; (X)$-type qubits at vertices $v_i,v_j\in V$ such that $v_i$ and $v_j$ are not neighbors and share no neighbors in common. Measuring $X_i\otimes X_j\; (Z_i\otimes Z_j)$ on these two qubits projects them into an effective two-dimensional subspace with the Pauli operators $\bar{X}=X_i\; (X_i\otimes X_j)$ and $\bar{Z}=Z_i\otimes Z_j\; (Z_i)$. As shown in Fig.~\ref{fig:TypeIFusionClusterState}(a), a new cluster state is obtained with vertices $v_i,v_j$ replaced by a single vertex $v_{ij}$ and an effective $Z\; (X)$-type qubit placed at this vertex. All the edges incident at $v_i$ and $v_j$ are incident at $v_{ij}$ in the new graph. To ensure that the new cluster state is the $+1$ eigenstate of all the stabilizers, a Pauli correction determined by the $X_i\otimes X_j\; (Z_i\otimes Z_j) $ measurement outcome, {{$m_{XX}=$ 0 or 1 ($m_{ZZ}=$ 0 or 1)}}, must be applied to the qubits that were originally adjacent to $v_j$. Specifically $Z^{m_{XX}}\; (Z^{m_{ZZ}})$ is applied to adjacent $X$-type qubits and $X^{m_{XX}}\; (X^{m_{ZZ}})$ is applied to adjacent $Z$-type qubits. 

With the above merging principle in mind, we introduce the 6-ring resource cluster state in Fig.~\ref{fig:TypeIFusionClusterState}(b). 
A copy of this state is placed at {two} opposite corners of each unit cell of the XZZX cluster state as shown in Fig.~\ref{fig:TypeIFusionClusterState}(c). Two qubits of the same type share a face centre or edge {centre}. If we measure $X\otimes X\; (Z\otimes Z)$ for each pair of $Z\; (X)$-type qubits sharing a face/edge, and apply the required Pauli corrections, we obtain the desired XZZX cluster state comprised of the effective qubits. Note that an unreliable $X\otimes X$ measurement on $Z$-type qubits only leads to an incorrect Pauli $Z$ correction to adjacent $X$-type qubits. Finally, we need to measure the effective Pauli $\bar{X}=X\otimes X$ of the effective $X$-type qubits and effective Pauli $\bar{Z}=Z\otimes Z$ of the effective $Z$-type qubits for error correction. Note that an unreliable $X\otimes X$ measurement in the second set of measurements is like a $\bar{Z}$ error on the effective $X$-type qubits. As before, Pauli corrections from the first set of measurements that create the cluster state may be simply tracked in software by re-interpreting the outcomes of the second set of measurements that are used for error correction. Overall then, error correction is implemented in our construction by measuring commuting observables $X\otimes X$ and $Z\otimes Z$, that is performing fusions, on pairs of qubits that share an edge or a face {centre}. The above discussion implies that biased fusion failure, as is the case in linear optics, effectively leads to biased Pauli $Z$ noise on the $X$-type qubits of the XZZX cluster state.

\begin{figure}
    \centering
    \includegraphics[width=\columnwidth]{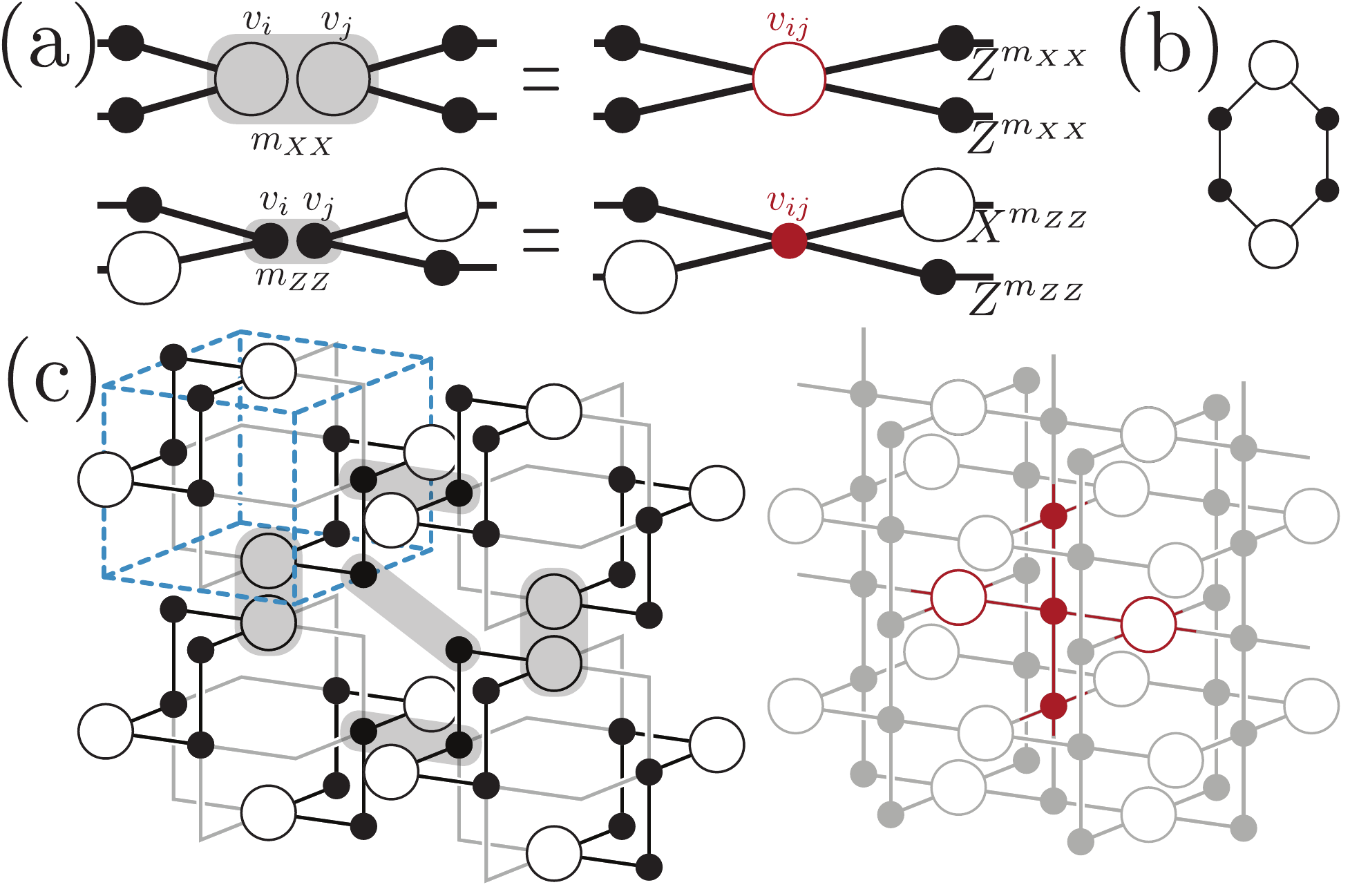}
    \caption{Building the XZZX cluster state using 6-ring resource states. (a) Measuring $X_i\otimes X_j\; (Z_i\otimes Z_j)$ on two $Z\; (X)$-type qubits projects those qubits onto a two-dimensional subspace with effective Pauli operators $\bar X_{ij} = X_i;\ (X_i\otimes X_j)$ and $\bar Z_{ij}=Z_i\otimes Z_j\; (Z_i)$. In terms of the effective Pauli operators, the two $Z\; (X)$-type qubits are replaced by a single $Z\; (X)$-type qubit $v_{ij}$, with all edges to neighbors intact. (b) The 6-ring resource state. (c) The fusion pattern used to join the 6-ring states to make the XZZX cluster state. }
    \label{fig:TypeIFusionClusterState}
\end{figure}

{\it Application to dual-rail qubits with linear-optic fusions---}A photonic dual-rail qubit is encoded in the presence of a single photon in one of two orthogonal modes, $\ket{\bar{0}}=\ket{01},\;\ket{\bar{1}}=\ket{10}$ and is a leading candidate for linear-optical quantum computing~\cite{knill2001scheme,nielsen2004optical,browne2005resource,kok2007linear}. In this platform, all single qubit gates, like the Hadamard gate, can be performed deterministically using passive linear optical elements~\cite{politi2009integrated}. Multi-qubit operations, like the fusion measurements, are non-deterministic. Fault-tolerant FBEC in this platform is based on heralded generation of few-body entangled resource states, followed by non-deterministic fusion measurements~\cite{li2015resource,bartolucci2021fusion}.
Remarkably, the resource states used in our proposal differ from the ones considered in previous works~\cite{bartolucci2021fusion} only by Hadamard transformations and these states can be easily generated using well-known linear-optics circuits. For clarity, in~\cite{supplement} we also give circuits for generation of our four-star and six-ring states. The required fusion measurements can be realized using a so-called type-II fusion circuit comprised of beamsplitters and photon number resolving detectors~\cite{browne2005resource,gilchrist2007efficient}. The circuit consumes the two dual-rail qubits which need to be fused and outputs the photon number/clicks observed at each detector. Conditional on the observed clicks, one of two operations is implemented by the circuit on the input qubits. Either a successful $X\otimes X$ and $Z\otimes Z$ measurement is implemented, in which case we say the fusion has succeeded and the measurement outcomes {are} inferred from the detector clicks. Or, each input qubit is independently measured in the $Z$-basis, in which case we say that the fusion has failed and the independent $Z$ measurement outcomes {are} again inferred from detectors clicks~\cite{bartolucci2021fusion,supplement}. Consequently, in a failed-fusion event, $Z\otimes Z$ can be recovered by multiplying the independent $Z$ measurement outcomes, while the $X\otimes X$ measurement is completely erased. Without any ancilla photons, the probability of fusion failure is $p_\mathrm{fail}=1/2$. With additional $(2^n-2)$-photon entangled ancilla, for a total of $2^n$ photons in the fusion circuit, the failure probability may be reduced to $p_\mathrm{fail}=1/2^n$~\cite{grice2011arbitrarily}. For the particular case of $n=2$, the failure probability may also be reduced to $1/4$ with a $4$-photon unentangled ancilla~\cite{ewert20143}. However, $n>2$ requires entangled states that get progressively more complicated to realize. 

So far we have ignored hardware imperfections that can introduce photon loss. Fortunately, the ideal fusion circuit preserves the number of photons, that is, the total detector clicks must be equal to the number of input photons. A loss of any photon in the fusion circuit will be heralded by observing fewer than expected clicks at the detector and result in an erasure of both $X\otimes X$ and $Z\otimes Z$ measurement outcomes. If the probability of loss per photon is $p_\mathrm{loss}$, then the probability of such an erasure in the boosted fusion circuit with a total of $1/p_\mathrm{fail}$ photons is $p_\mathrm{full\; erase}=1-(1-p_\mathrm{loss})^{1/p_\mathrm{fail}}$~\cite{bartolucci2021fusion}. This equation highlights the tradeoff between the rate at which only the $X\otimes X$ outcome is erased due to fusion-failure and the rate at which both $X\otimes X$ and $Z\otimes Z$ outcomes are erased due to photon loss. As we decrease $p_\mathrm{fail}$ by adding more photons to the fusion circuit, we increase the probability of photon loss from the fusion circuit $p_\mathrm{full\; erase}$.

{W}e evaluate the performance of our fusion architectures under the linear optical error model including fusion failures and photon loss as described above. We perform Monte-Carlo simulations of errors when the four-star and six-ring states are used for FBEC with $d\times d\times d$ XZZX cluster states. The syndrome data generated by the errors is arranged on a graph and decoded using the linear-time erasure-decoder~\cite{delfosse2020linear}. We evaluate the logical error rates for different cluster state sizes $d$, and error parameters $p_{\mathrm{fail}}$, and $p_\mathrm{loss}$ and estimate the threshold which is plotted in Fig.~\ref{fig:FailVsErasePlot}. The solid red and blue curves are the thresholds for the 4-star and 6-ring constructions respectively. If $p_\mathrm{loss}$ and $p_\mathrm{fail}$ lie under the curves, then these errors are correctable, otherwise they are not. {{We also compare our results with the thresholds obtained with (a) the 4-star and 6-ring construction from a previous work~\cite{bartolucci2021fusion} to construct the XZZX cluster state and (b) our 4-star and 6-ring constructions based on the adaptive error-correction strategy introduced in~\cite{auger2018fault}. The thresholds with both (a) and (b), shown in Fig.~\ref{fig:FailVsErasePlot} using dashed lines, are identical to each other as they have identical error syndrome graphs and are consistent with known results~\cite{bartolucci2021fusion}.}}

{We} observe, {first}, that the thresholds for the six-ring construction are higher than the four-star construction as it has fewer fusions, and hence a lower probability of error, per cluster state qubit. In the absence of photon loss, the numerically obtained threshold for biased fusion-failure using our scheme is $34.7\%$ for the six-ring construction and $20.6\%$ for the four-star construction. {Without photon loss, these thresholds can also be derived analytically (see Supplement~\cite{supplement})} and are significantly higher than the threshold for fusion-failure obtained with previous proposals, which are correspondingly $\sim 24\%$ for the six-ring construction ~\cite{bartolucci2021fusion} and $\sim14.5\%$ for the four-star construction~\cite{auger2018fault}, {{that fail to leverage the noise bias in fusion failures}}. This is because, as argued earlier, in our approach fusion failure leads to two-dimensional system symmetry and hence a two-dimensional syndrome graph which is easier to decode. In contrast, {in} previous strategies the syndrome graph resulting from fusion failures is three-dimensional, which is harder to decode.  

When $p_\mathrm{loss}\neq 0$, we must deal with both full fusion erasure along with fusion failure. Increase in $p_\mathrm{full\; erase}$ means we must decrease $p_\mathrm{fail}$ by using additional photons. However, this also increases $p_\mathrm{full\; erase}$ due to the possibility of these additional photons being lost. This relationship  between $p_\mathrm{full\; erase}$, $p_\mathrm{fail}$, and $p_\mathrm{loss}$ gives the overall ``inverted-u" shape of the threshold curve. Importantly, from Fig.~\ref{fig:FailVsErasePlot} we see that our scheme with the six-ring resource states can tolerate up to $ 0.37\%$ of photon loss with $25\%$ fusion failure. That is, it is possible to achieve fault tolerance with only $2$ additional entangled ancilla photons for boosted fusions~\cite{grice2011arbitrarily}. More importantly, this means that we could even reach fault tolerance with boosted fusions using only $4$ additional unentangled ancilla photons~\cite{ewert20143} and { {$p_\mathrm{loss}\leq 0.25\%$}}. On the other hand, the maximum fusion failure that previous schemes can tolerate is $< 25\%$, meaning they could not achieve fault-tolerance with only $2$ additional entangled ancilla photons or $4$ additional unentangled ancilla photons.

\begin{figure}
    \centering
    \includegraphics[width=\columnwidth]{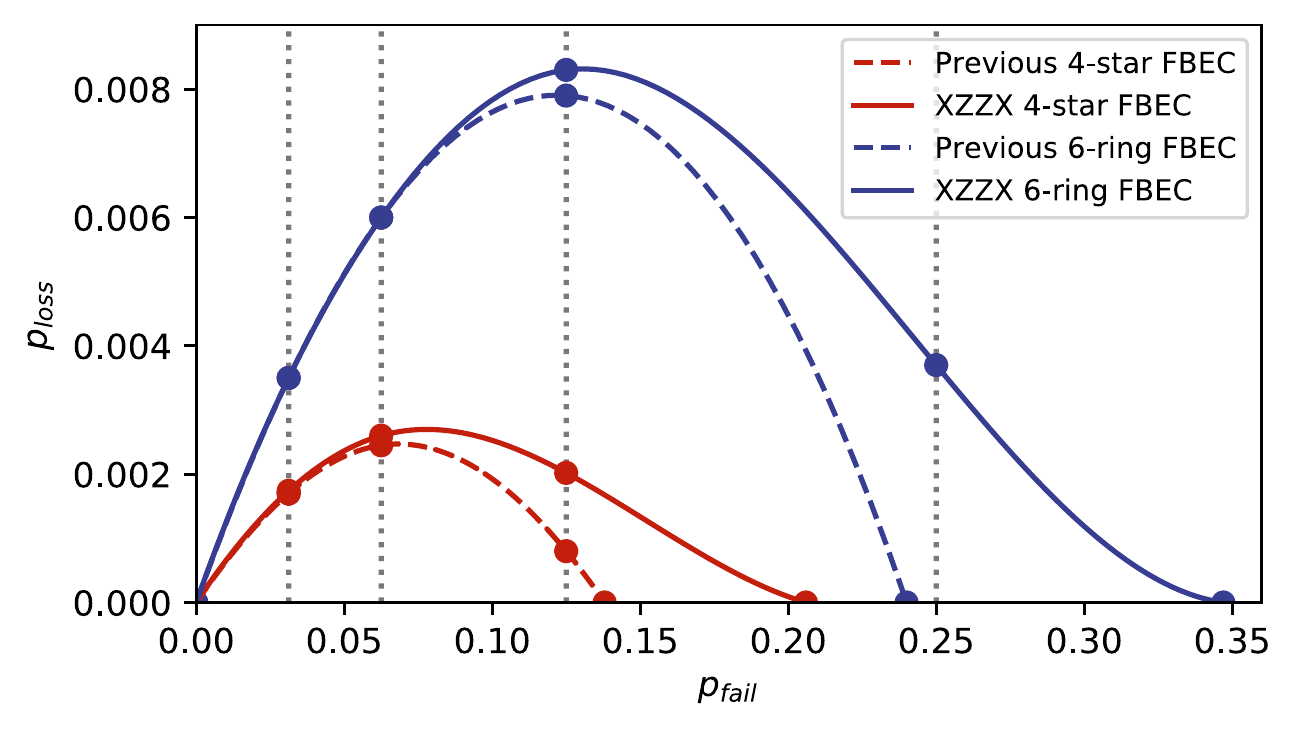}
    \caption{{{Threshold with four-star (red) and six-ring (blue) construction protocols as proposed here under the linear optical error model. For comparison, thresholds based on previous approaches which do not leverage the noise bias in fusion failures is also shown~\cite{bartolucci2021fusion, auger2018fault}.}} Excluding XZZX FBEC points on the x-axis, all  points are simulated thresholds for sets of cluster states of size $d\times d \times d$, with $d$ up to $17$. XZZX FBEC points on the x-axis are thresholds for sets of $d \times d \times 4$ cluster states, with $d \leq 71$. From left to right, the black dotted vertical lines represent fusion success boosting by 30, 14, 6, and 2 additional entangled ancillae photons. }
    \label{fig:FailVsErasePlot}
\end{figure}

{\it Conclusion} By taking advantage of the biased structure of fusion failures, we have introduced new resource states and fusion strategies for FBEC that allow for more efficient error correction of biased fusion failures. This FBEC strategy is particularly relevant to linear-optical quantum computers based on dual-rail photonic qubits, where biased fusion failures are the dominant source of error. Our resource states and fusion strategies require no additional overhead to realize compared to the previous approach of Ref.~\cite{bartolucci2021fusion}, but result in higher thresholds to fusion failures for both the 4-star and 6-ring resource states. In the 6-ring construction in particular, our strategy has a threshold over $25\%$ to fusion failures, which can be reached using only a $2$-photon entangled ancilla or a $4$-photon unentangled ancilla; thus, our construction overcomes a key barrier for photonic quantum computing.

Our construction achieves higher thresholds because: (1) the fusion operations in linear optics have a natural bias towards $Z$ errors, (2) we construct our cluster state in a bias-preserving way so that the final cluster state also has a bias towards $Z$ errors, and (3) XZZX cluster state is designed to correct $Z$ errors~\cite{claes2022tailored}. A similar strategy should apply to any other hardware with a similar bias in the entangling operations. Investigating other classes of hardware that can benefit from our overall strategy will be left to future work.

\begin{acknowledgments}
{\it Acknowledgements--} This material is based on work supported by the National Science Foundation (NSF) under Award No. 2137740. Any opinions, findings, and conclusions or recommendations expressed in this publication are those of the authors and do not necessarily reflect the views of NSF.
\end{acknowledgments}

\putbib[references]
\end{bibunit}


\setcounter{equation}{0}
\setcounter{figure}{0}
\setcounter{table}{0}
\setcounter{section}{0}
\makeatletter

\renewcommand{\theequation}{S\arabic{equation}}
\renewcommand{\thefigure}{S\arabic{figure}}
\renewcommand{\thetable}{S\arabic{table}}
\renewcommand{\thesection}{S\arabic{section}}

\title{Supplemental Material: Tailoring fusion-based error correction for high thresholds to biased fusion failures}

\clearpage
\maketitle
\onecolumngrid

\begin{bibunit}[apsrev_longbib]

\section{4-star construction} 
\begin{figure}[ht]
    \centering    \includegraphics[width=\columnwidth]{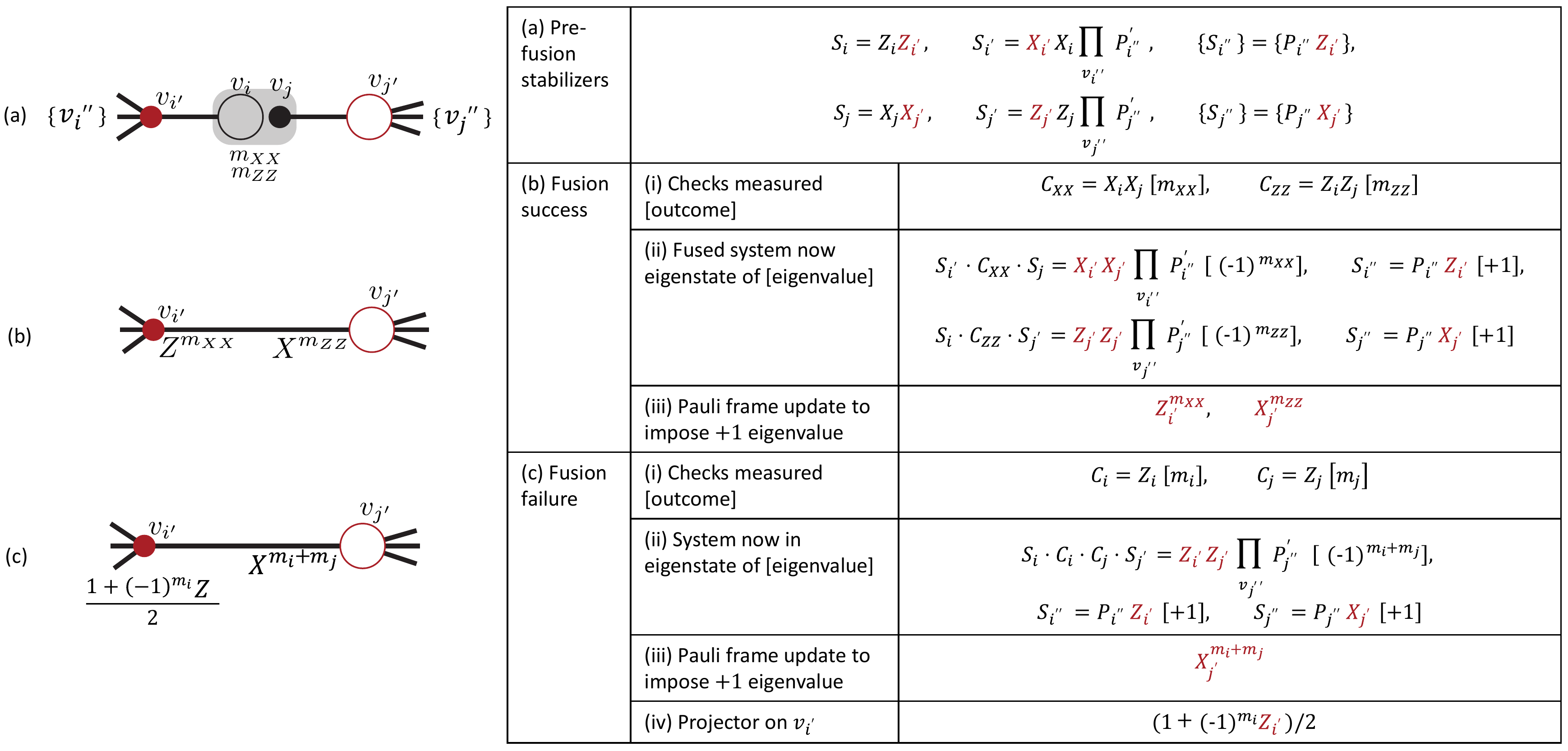}
    \caption{Tracking the stabilizers and deriving the required Pauli frame updates for a fusion conducted on qubits at vertices $v_i,v_j$ when $v_{i'}\in\mathcal{X}$ and $v_{j'}\in\mathcal{Z}$. We look at the state of the system (a) pre-fusion, (b) post-successful fusion, and (c) on fusion failure. Note that $P_{i^{''}}$ ($P_{j^{''}}$) denotes a Pauli operator on $v_{i^{''}}$ ($v_{j^{''}}$).}
\label{fig:4starstabilizercheckXZ}
\end{figure}

\begin{figure}[ht]
    \centering    \includegraphics[width=\columnwidth]{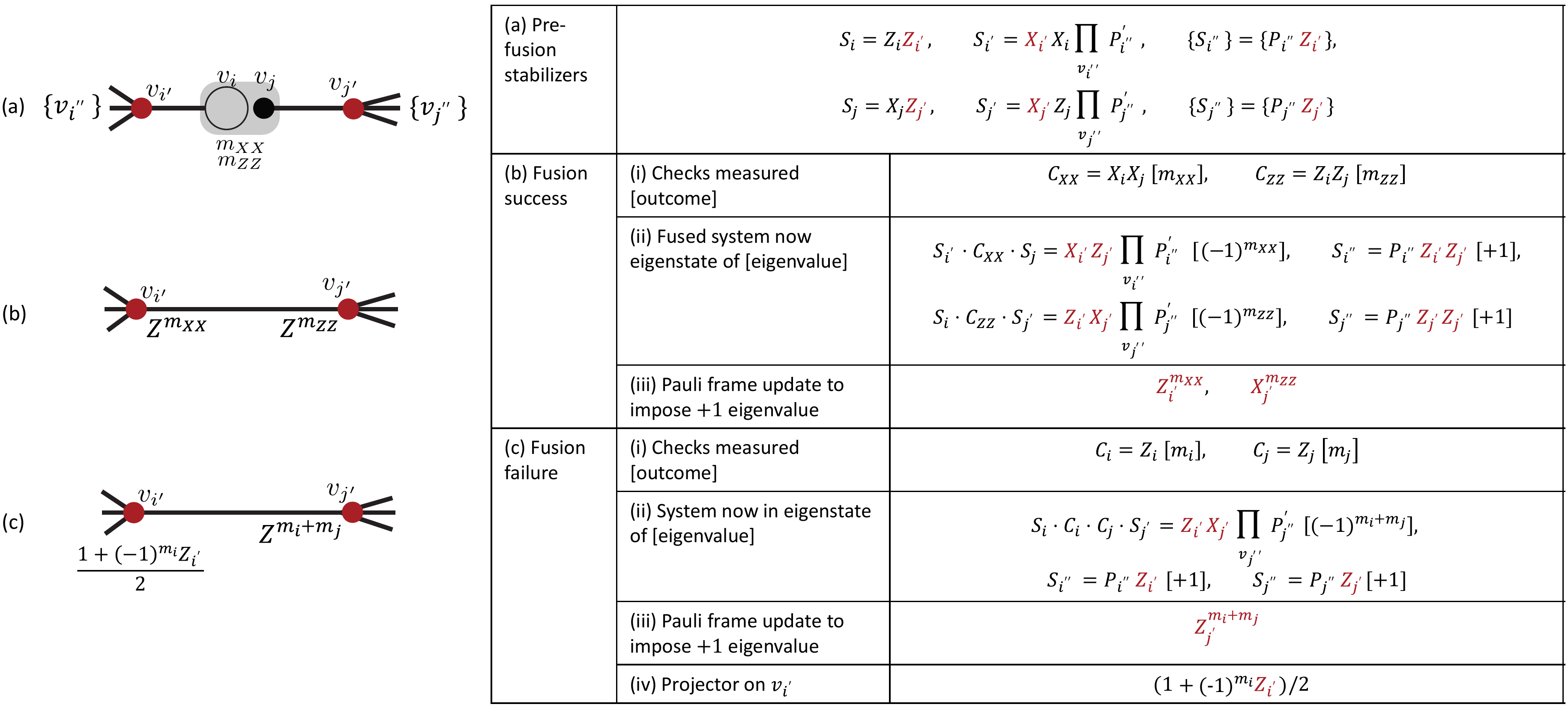}
    \caption{Tracking the stabilizers and deriving the required Pauli frame updates for a fusion conducted on $v_i,v_j$ when $v_{i'},v_{j'}\in\mathcal{X}$. We look at the state of the system (a) pre-fusion, (b) post-successful fusion, and (c) on fusion failure. }
\label{fig:4starstabilizercheckXX}
\end{figure}

Here we show the evolution of stabilizers when a fusion operation is conducted between two dangling qubits in the XZZX cluster state. 
Continuing with the notation in the main text, we use a cluster state defined on a graph $G$. Let $G$ have a  dangling qubit at vertex $v_i \in \mathcal{Z}$ and $v_j \in \mathcal{X}$, each with an edge to a qubit at vertex $v_{i'}$ and $v_{j'}\neq v_{i'}$ respectively. Qubit $v_{i'}$ ($v_{j'}$) has further edges to a set of qubits $\{v_{i^{''}}\}$ ($\{v_{j^{''}}\}$). In Fig.~\ref{fig:4starstabilizercheckXZ}, we show the evolution of stabilizers and associated Pauli frame updates when  $v_{i'}\in\mathcal{X}$ and $v_{j'}\in\mathcal{Z}$. In Fig.~\ref{fig:4starstabilizercheckXX} we show the same for  $v_{i'},v_{j'}\in\mathcal{X}$. The set of peripheral qubits $\{v_{i^{''}}\}$ 
 and $\{v_{j^{''}}\}$  are not shown for simplicity, and indeed we shall see that they are not affected by the fusion. Note that after the fusion, the qubits at vertices $v_i, v_j$ are disentangled from the cluster state.
\\
Pre-fusion, the stabilizer generators $\{S_r\}$ are defined according to Eq.~\eqref{eq:GeneralizedClusterStabilizers} in the main text.  If the fusion succeeds (fails), we measure checks $X_i \otimes X_j, Z_i\otimes  Z_j$ ($Z_i , Z_j$) with measurement outcomes $m_{XX}, m_{ZZ} $ ($m_i, m_j$). Post-fusion, the system is constrained by the surviving stabilizer set generated by (i) the original stabilizers that commute with the measurements and (ii) the commuting products of those stabilizers that do not. The system will now be in the eigenstate of this new generating set, given in row b(ii) of the table in Fig.~\ref{fig:4starstabilizercheckXZ},\ref{fig:4starstabilizercheckXX}, with eigenvalues defined by the measurement outcomes. Thus, in order to restore the system to the $+1$ eigenstate of  a given stabilizer generator $S_p$, we apply a Pauli frame update that is identified by locating the correction that anticommutes with $S_p$ but no other stabilizer generator. These corrections are shown in row b(iii). In practice, this Pauli correction is tracked in software. 

In the case when a fusion failure occurs, the stabilizer of the combined system centered at $v_{j^{\prime}}$ is recovered by adding the fusion measurement outcomes. In particular, when $v_{j^\prime} \in \mathcal{Z}$ ($v_{j^\prime} \in \mathcal{X}$), the conditional Pauli frame update applied to it is an $X$ ($Z$) operator. However, a projector $(1 + (-1)^{m_i} Z) / 2$ is applied to vertex $v_{i^{\prime}}$, which destroys its X measurement outcome information. This is equivalent to a successful fusion followed by a $Z$ measurement on $v_{i'}$, a process described by the error channel ${\rho\mapsto}(\rho+Z_{i'}\rho Z_{i'})/2$.

Additionally, we note that the sets of stabilizers $\{S_{i^{''}} \}, \{S_{j^{''}} \}$ centered at the neighbouring qubits remain undisturbed in all cases. We shall use this fact to simplify our analysis in the next section.

\section{6-ring construction} 
\begin{figure}[ht]
    \centering
\includegraphics[width=\columnwidth]{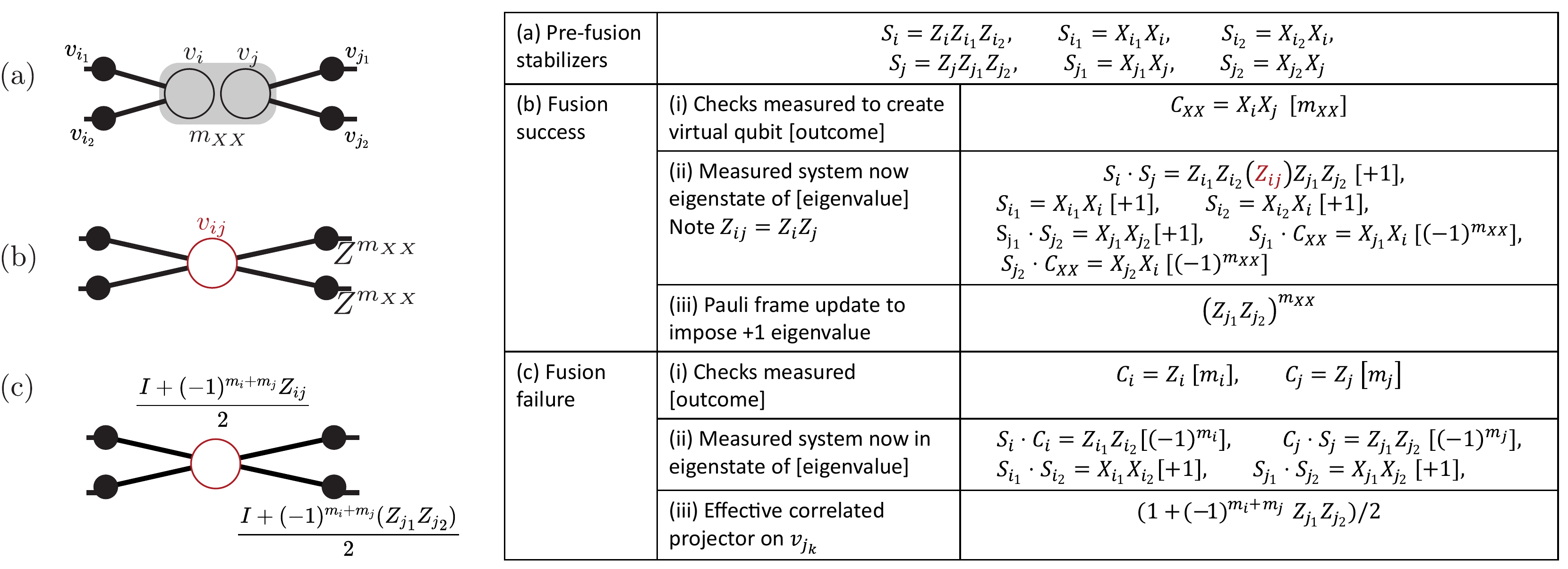}
    \caption{Tracking the stabilizers and deriving the required Pauli frame updates for the six-ring construction. Measurements are conducted on $v_i,v_j$ with $v_{i'},v_{j'}\in\mathcal{Z}$. We look at the system (a) pre-fusion, (b) post-successful fusion, and (c) on fusion failure.}
    \label{fig:6ringstabzz}
\end{figure}

In this section we show how measuring $X\otimes X$ $(Z\otimes Z)$ for $Z$-type ($X$-type) qubits in six-ring cluster states produces virtual qubits with Pauli corrections on neighboring qubits and derive the noise introduced in the cluster state due to erasure of $X\otimes X$ measurement outcome. Let the cluster state graph $G$ have a degree-2 vertex $v_i$ ($v_j$ ) with an edge to two qubits at vertices $\{v_{i_k}\}$ {($\{v_{j_k}\}$) with $\{v_{j_k}\}\cap\{v_{j_k}\}=\emptyset$}. An effective cluster state qubit $v_{ij}$ is formed when such a measurement is performed between $v_i$ and $v_j$. In Fig.~\ref{fig:6ringstabzz}, we show the evolution of stabilizers and associated Pauli frame updates when $v_{i}, v_{j}\in\mathcal{Z}$. Note that we have disregarded further qubits connected to $\{v_{i_k}\}$, and $\{v_{j_k}\}$ since their stabilizers remain unchanged during this process, just like in the 4-star construction discussed in the previous section.

Similar to the previous section,  we update the stabilizers according to the checks measured. Clearly we see that measuring $X_i\otimes X_j$ creates a virtual qubit with effective Pauli $\bar Z=Z_i\otimes Z_j$. 
Note that on fusion failure, due to the form of the system stabilizers in this setup, two X-type qubits initially connected to $v_j$ experience a correlated projection into an eigenstate of $Z_{j_1} Z_{j_2}$. Individual errors on either of the two qubits $v_{j_1}, v_{j_2}$ connect unit cell syndromes in mutually perpendicular directions in the cluster state. As a result, this correlated error effectively connects a unit cell syndrome with its diagonally displaced neighbour.

\section{4-star resource state generation}
\begin{figure}[ht]
    \centering
\includegraphics[width=\columnwidth]{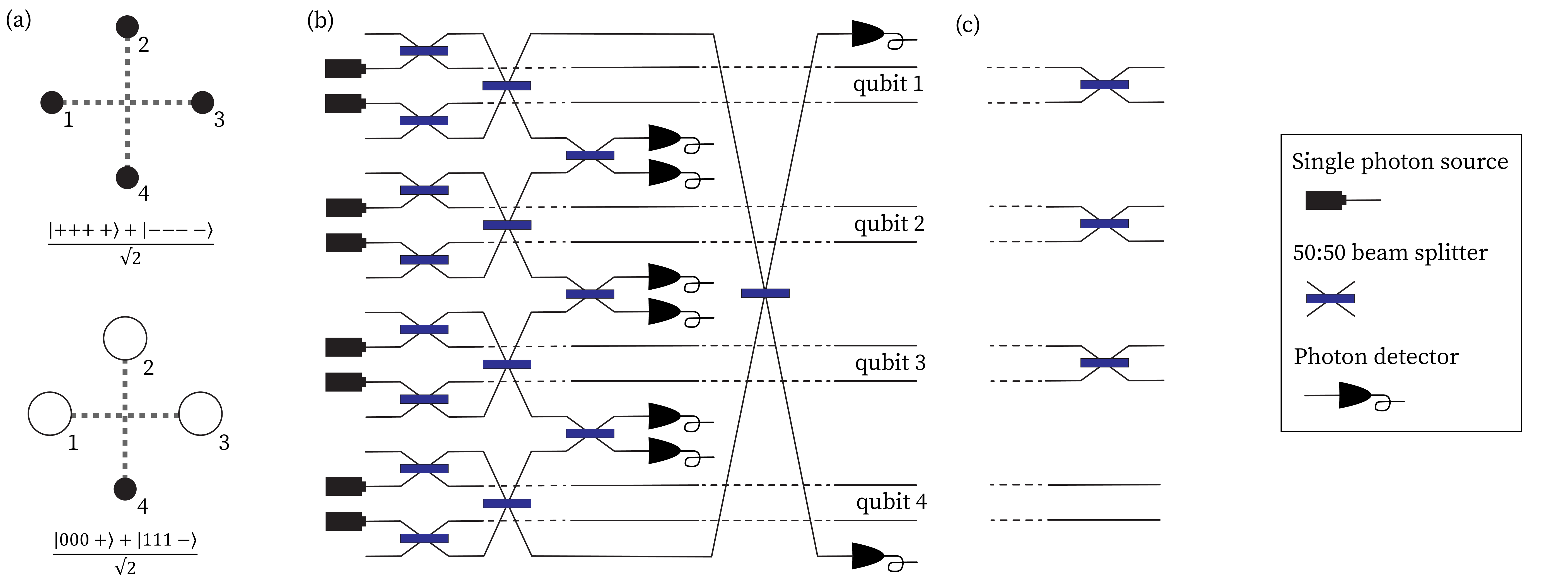}
    \caption{Construction of the four-star resource state for the XZZX cluster state. (a) Representations of the two resource states (b) Probabilistic photonic fusion circuit to construct the state  $\left(|++++\rangle + |----\rangle \right)/ \sqrt{2}$. (c) Beam splitters are applied to three qubits in the state output from (b) to obtain the second resource state $\left(|000+\rangle + |111-\rangle \right)/ \sqrt{2}$.}
    \label{fig:4starcircuit}
\end{figure}

Four-star resources states can probabilistically be generated from single photons by a sequence of beam splitters and photon detectors, conditioned on a certain detector output. A potential generation circuit is shown in Fig.~\ref{fig:4starcircuit}, based on prior constructions for three-body GHZ states \cite{varnava2008good, li2015resource}. This circuit succeeds if a single photon is output at each detector pair, where a detector pair is defined at the two output ports of a single beam splitter.

\section{6-ring resource state generation}
\begin{figure}[ht]
    \centering
    \includegraphics[width=\columnwidth]{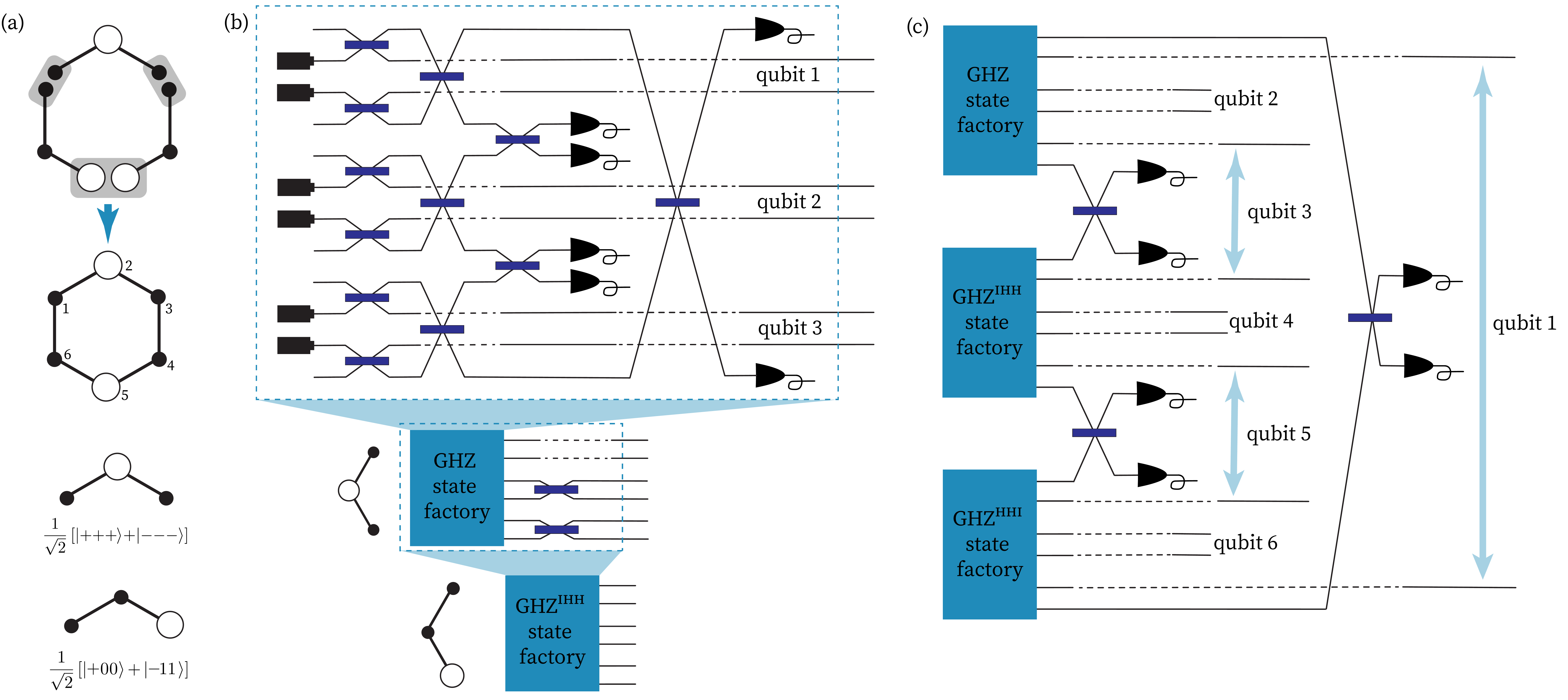}
    \caption{Construction of the 6-ring resource state for the XZZX cluster state. (a) (top) Type-I fusions are conducted between input GHZ states to form the resource state. (bottom) Representations of the input GHZ states. (b) Probabilistic photonic circuit to construct the state $\left(|+++\rangle + |---\rangle \right)/ \sqrt{2}$ \cite{li2015resource}. Hadamards can be applied to specific qubits to obtain the full input state set. (c) Photonic circuit to conduct Type-I fusions between output GHZ states to construct a six-ring state.}
    \label{fig:6ringcircuit}
\end{figure}

As described in Fig.~\ref{fig:6ringcircuit}(a), a six-ring resource state can be  constructed by performing Type-I fusions \cite{browne2005resource} on a set of GHZ states. We present a potential photonic circuit to probabilistically construct six-rings from single photons in Fig.~\ref{fig:6ringcircuit}(b,c).  The base GHZ state 
 $\left(|+++\rangle + |---\rangle \right)/ \sqrt{2}$ (upto heralded Pauli corrections) is generated by the  circuit shown in  Fig.~\ref{fig:6ringcircuit}(b) when one photon is detected at each pair of detectors. This occurs with probability $1/32$, which can be effectively increased to near-unity via multiplexing. Two beam splitters performing Hadamard operations to specific qubits can be applied to this GHZ state in order to get the complete input set of GHZ states. These input states are fed into the circuit in  Fig.~\ref{fig:6ringcircuit}(c), which performs Type-I fusions between the peripheral qubits of the individual GHZ states. Note that this secondary circuit layer has a success probability $(1/2)^3 $ which can be increased by further multiplexing. Derivation of a more efficient circuit is left to future work.

\section{Biased Fusion Failures in Boosted Type-II Fusion}
The aim of the fusion measurements is to measure $X_i\otimes X_j$ and $Z_i\otimes  Z_j$ between two qubits labelled by $i,j$. That is, in terms of the dual rail encoding $\ket{\bar{0}}={ \ket{10}},\; \ket{\bar{1}}=  { \ket{01}}$, the aim is to discriminate between the four states $\ket{\psi_\pm}=(\ket{\bar{0}_i\bar{1}_j}\pm \ket{\bar{1}_i\bar{0}_j})/\sqrt{2}$ and $\ket{\phi_\pm}=(\ket{\bar{0}_i\bar{0}_j}\pm \ket{\bar{1}_i\bar{1}_j})/\sqrt{2}$ which are the simultaneous 
eigenstates of $X_i X_j$ and $Z_i Z_j$:
\begin{align}
Z_iZ_j\ket{\psi_\pm}=-\ket{\psi_\pm},\; X_iX_j\ket{\psi_\pm}=\pm\ket{\psi_\pm},\;
Z_iZ_j\ket{\phi_\pm}=\ket{\phi_\pm},\; X_iX_j\ket{\phi_\pm}=\pm\ket{\phi_\pm}
\end{align}
\begin{figure}[ht]
    \centering
\includegraphics[width=0.16\columnwidth]{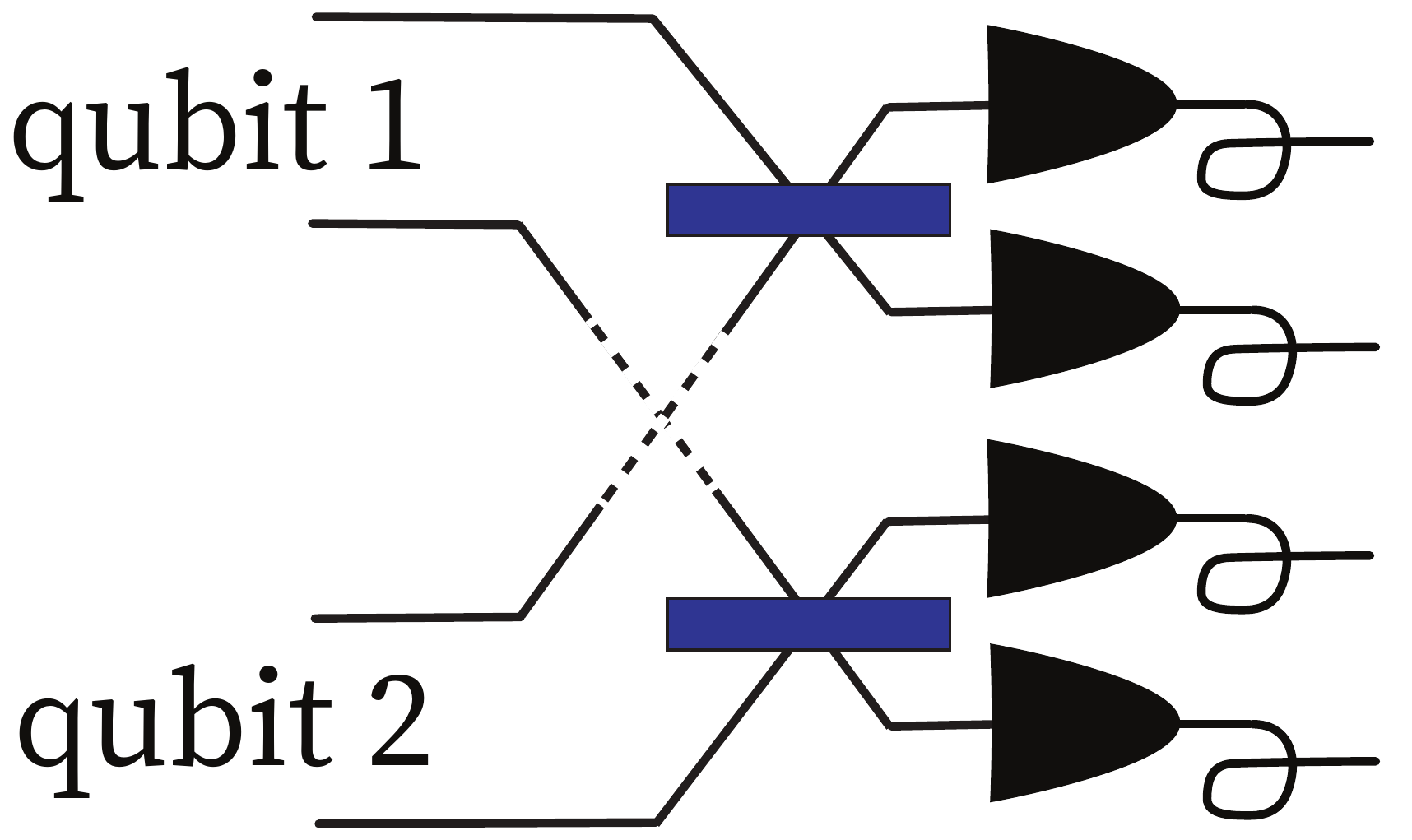}
    \caption{Photonic circuit for type-II fusion between two dual rail qubits using beam splitters and photon detectors.}
    \label{fig:type2photonic}
\end{figure}
In a direct type-II fusion~\cite{browne2005resource,gilchrist2007efficient}, the four modes of the two dual-rail qubits are interfered using two $50:50$ beam-splitters as shown in Fig.~\ref{fig:type2photonic}. The transformed states at the ouput of the network are:
\begin{equation}
\begin{aligned}
\ket{\psi_+}&\rightarrow \frac{i}{\sqrt{2}}(\ket{1100}+\ket{0011})\\
\ket{\psi_-}&\rightarrow \frac{i}{\sqrt{2}}(\ket{1001}-\ket{0110})\\
\ket{\phi_+}&\rightarrow \frac{i}{{2}}(\ket{2000}+\ket{0200}+\ket{0020}+\ket{0002})\\
\ket{\phi_-}&\rightarrow \frac{i}{{2}}(\ket{2000}-\ket{0200}+\ket{0020}-\ket{0002})
\end{aligned}
\end{equation}
As can be seen, the photon number distribution $n$ across the four output modes, which can be measured using PNRDs, carries information about the input states. The output photon number distribution generated if $\ket{\psi_+}$ is input into the interference network, $N_{\ket{\psi_+}}=\{1100\}$ or $\{0011\}$, discriminates $\ket{\psi_+}$ from other states. Similarly, the output photon number distribution generated if $\ket{\psi_-}$ is input into the interference network, $N_{\ket{\psi_-}}=\{1001\}$ or $\{0110\}$, discriminates $\ket{\psi_-}$ from other states. Unfortunately, however, it is not possible to discriminate between $\ket{\phi_+}$ and $\ket{\phi_-}$ using PNRDs as they produce the same output photon number distribution: $N=\{2000\}$ or $N=\{0200\}$ or $N=\{0020\}$ or $N=\{0002\}$. In this case we say that the fusion fails. Given an equal distribution of Bell states at the input (which is the case in 4-star and 6-ring fusions), the probability of failure is $50\%$. However, note that measuring $N=\{2000\}$ or $N=\{0020\}$ discriminates $(\ket{\phi_+}+\ket{\phi_-})/\sqrt{2}=\ket{0_i0_j}$ and the output photon number distribution $N=\{0200\}$ or $N=\{0002\}$ discriminates $(\ket{\phi_+}-\ket{\phi_-})/\sqrt{2}=\ket{1_i1_j}$. Thus, even when fusion fails, $Z_i$ and $Z_j$ of each dual-rail qubit is still measured. 

Refs.~\cite{grice2011arbitrarily,ewert20143} introduced protocols for fusions with success probability boosted by lifting the degeneracy in the output photon-number distribution when $\ket{\phi_\pm}$ are input into a interference network. Here, we give a broad overview of the two protocols without repeating the details in Refs.~\cite{grice2011arbitrarily,ewert20143}. The broad overview will be sufficient to see that even when boosted fusions fail, $Z_i$ and $Z_j$ of each dual-rail qubit is still measured.

The protocols in~\cite{grice2011arbitrarily,ewert20143} use auxillary entangled-photons, $\ket{\Gamma}$. The exact form of this entangled state is not important for our purpose here, but $\ket{\Gamma}$ required for the protocol in ~\cite{grice2011arbitrarily} is different from that in ~\cite{ewert20143}. $\ket{\Gamma}$, together with the state of the two duil-rail qubits, transform through the interference network as follows,
\begin{equation}
   \begin{aligned}
\ket{\psi_+}\ket{\Gamma}&\rightarrow |\ket{N'_{\psi_+}}\\
\ket{\psi_-}\ket{\Gamma}&\rightarrow |\ket{N'_{\psi_-}}\\
\ket{\phi_+}\ket{\Gamma}&\rightarrow x|\ket{N'_{\phi_+}}+y(\ket{A}+\ket{B})\\
\ket{\phi_-}\ket{\Gamma}&\rightarrow x|\ket{N'_{\phi_-}}+y(\ket{A}-\ket{B})
\end{aligned} 
\end{equation}
Here, $x,y$ are c-numbers for normalization and $|\ket{N'_{\psi_\pm}}$, $|\ket{N'_{\phi_\pm}}$, $\ket{A}$, $\ket{B}$ are states with distinct photon-number distributions. $|\ket{N'_{\psi_\pm}}$, $|\ket{N'_{\phi_\pm}}$ may be a superposition of many Fock states (indicated by $|\ket{\;}$). Clearly, the protocol uniquely identifies $\ket{\psi_\pm,\phi_\pm}$ if PNRDs measure photon number distributions consistent with  $|\ket{N'_{\psi_\pm}},\;|\ket{N'_{\phi_\pm}}$. However, the protocol fails if photon number distribution corresponding to the states $\ket{A}$ or $\ket{B}$ is measured. Measuring photon number distribution consistent with $\ket{A}$ or $\ket{B}$ discriminates $(\ket{\phi_+}+\ket{\phi_-})/\sqrt{2}$ or $(\ket{\phi_+}-\ket{\phi_-})/\sqrt{2}$ respectively and, like the direct type-II measurement, this is equivalent to measuring $Z_i$, $Z_j$. This feature of fusion measurements has also been identified in ~\cite{bartolucci2021fusion}. The key difference in the protocols of ~\cite{grice2011arbitrarily} and ~\cite{ewert20143} is the nature of $\ket{\Gamma}$ and the success probability. In ~\cite{grice2011arbitrarily} a success probability of $(1-2^{-n})$ is obtained with $(2^n-2)$-photon entangled state, while in ~\cite{ewert20143}, twice as many photons are required to reach the same success probability. Importantly, with the protocol of~\cite{ewert20143}, $\ket{\Gamma}$ for $75\%$ successful fusion is an unentangled state of $4$ photons.

\section{Analytic threshold in case of only biased fusion failures}

Consider the four-star construction from Fig.~\ref{fig:TypeIIFusionClusterState} in the main text. Suppose the probability of erasing only $X\otimes X$ measurement outcome or the probability of a biased fusion failure is $p$. An incorrect $X\otimes X$ measurement implies that an incorrect Pauli $Z$ correction may be applied to a $X$ type qubit that eventually makes the cluster state. Each such $X$ type qubit gets a Pauli $Z$ correction from the fusions of three neighboring dangling qubits. Thus, effectively the probability of a $Z$ error on each $X$ type qubit forming the cluster state is $3p(1-p)^2+3p^2(1-p)+p^3$. The syndrome graph in 2D is the $4^4$ tiling of the plane with a bond percolation threshold of $50\%$~\cite{stace2009thresholds}. Thus, the threshold $p_\mathrm{th}$ may be found by requiring that $3p_\mathrm{th}(1-p_\mathrm{th})^2+3p_\mathrm{th}^2(1-p_\mathrm{th})+p_\mathrm{th}^3=0.5$ or $p_\mathrm{th} = 20.63\%$.

In order to justify the threshold for biased fusion failures with the six-ring construction, we examine the syndrome graph formed by fusion failures between resource states. A fusion failure between two $X$-type qubits creates a an erasure on this qubit, leading to $(-1)$ stabilizer outcomes on adjoining unit cells within a plane. When a fusion between two $Z$-type qubits fails, the two adjacent $X$-type qubits in the resource state are projected into an eigenstate of $Z \otimes Z$. This leads to a pair of $(-1)$ stabilizer outcomes on unit cells diagonally displaced from one another in the same plane. Note that all such diagonals have the same orientation. As a result, the syndrome graph in 2D results in a $3^6$ tiling of the plane. This has a percolation threshold of $2\sin(\pi/18)\approx 34.73\%$ \cite{sykes1964exact}, which agrees with numerical simulations of this construction at large lattice sizes.

\putbib[references]
\end{bibunit}

\end{document}